\documentstyle[aps,prl,epsfig]{revtex}
\tightenlines
\begin{document}
\twocolumn[\hsize\textwidth\columnwidth\hsize\csname@twocolumnfalse\endcsname
\draft
\title{A new monte carlo algorithm for growing compact self avoiding walks}
\author{S.L. Narasimhan$^{1*}$,  P.S.R. Krishna$^{1*}$, K.P.N.Murthy$^{2+}$ and M. Ramanadham$^{1\dagger}$}
\address{$^1$Solid State Physics Division,
\\ Bhabha Atomic Research Centre, Mumbai - 400 085, India}
\address{$^2$Materials  Science  Division, Indira Gandhi Centre for Atomic Research,
\\ Kalpakkam - 603 102, Tamilnadu, India}

\maketitle

\begin{abstract}
We   propose  an  algorithm  based  on  local  growth  rules  for
kinetically generating self avoiding walk configurations  at  any
given  temperature. This algorithm, called the Interacting Growth
Walk (IGW) algorithm, does not suffer from attrition on a  square
lattice   at   zero  temperature,  in  cotrast  to  the  existing
algorithms.  More  importantly,  the  IGW  algorithm  facilitates
growing  compact configurations at lower temperatures - a feature
that makes it attractive for studying a variety of processes such
as the folding of proteins. We  demonstrate  that  our  algorithm
correctly  describes  the collapse transition of a homopolymer in
two dimensions.
\end{abstract}
\vfill
\twocolumn
\vskip.5pc]
\narrowtext
The  configurational  properties  of linear polymers undergoing a
collapse transition at a  tricritical  temperature  $T_{\theta}$,
called  the $\theta$-point, have been studied extensively because
of their relevance to a wide variety of applications such as, for
example, the protein folding problem [1]. The average  radius  of
gyration  (or  equivalently, the average end-to-end distance) and
the configurational entropy of a long polymer chain have  a  {\it
universal}  (i.e., system-independent) behaviour characterized by
the  exponents  $\nu$  and  $\gamma$  respectively   [2].   These
exponents  have distinct sets of values for the three temperature
regimes, $T > T_{\theta}$, $T = T_{\theta}$ and $T <  T_{\theta}$
[2,3].  In  order to understand the statistical nature of polymer
conformations in these three universal regimes, Interacting  Self
Avoiding  Walk  (ISAW) models with appropriate non-bonded nearest
neighbour (nbNN) interactions have been proposed [4].

Let  ${\cal  S}_{N}$  denote  an  ensemble  of  equally  weighted
$N$-step SAW configurations, generated on a lattice by a standard
algorithm[5]. If $\epsilon_{0}$ is the energy associated with any
nbNN contact, a SAW configuration with a total of  $n_{NN}$  such
contacts will have an energy $E = n_{NN}\epsilon_{0}$. Hence, one
can  assign  to  it a Boltzmann weight proportional to $e^{-\beta
E}$, where $\beta = 1/k_{B}T$, $k_{B}$ is the Boltzmann  constant
and   $T$   the   temperature.   Such   Boltzmann   weighted  SAW
configurations constitute an ISAW  ensemble,  denoted  by  ${\cal
I}_{N}(\beta)$.  By  this  definition, ${\cal I}_{N}(\beta=0)$ is
the same as ${\cal S}_{N}$ because all the configurations of  the
former  have  the  same probability of occurrence irrespective of
their energies. Therefore, in the context of the  ISAW  ensemble,
${\cal  S}_{N}$  can  be  thought of as representing a polymer at
'infinite' temperature. The statistical accuracy of any  physical
quantity  averaged  over  ${\cal I}_{N}(\beta)$ becomes poorer at
lower temperatures because significant contribution comes from  a
smaller  number  of  configurations  [6]. In order to improve the
statistics, especially at low temperatures, it  is  necessary  to
generate  as large an ensemble, ${\cal S}_{N}$, as possible; this
process could become prohibitively slow due to  severe  attrition
for large $N$.

A  better  solution  is  to devise an algorithm based on suitable
geometrical ({\it athermal}, or 'infinite' temperature) rules for
generating an ensemble, ${\cal G}_{N}$, identically equivalent to
${\cal I}_{N}(\beta > 0)$. For example, the Kinetic  Growth  Walk
(KGW)  [7]  or  the  Smart  Kinetic Walk (SKW) [8] on a hexagonal
lattice straightaway  generates  an  ensemble  of  configurations
equivalent  to  the  ISAW  ensemble, ${\cal I}_{N}(\beta = ln2)$.
Having generated the {\it athermal} ensemble, ${\cal G}_{N}$,  by
such  a geometric algorithm, ensemble averages corresponding to a
lower temperature could be obtained by Botzmann  weighting  these
configurations    appropriately.   This   would   ensure   better
statistical accuracy  as  compared  to  what  could  be  obtained
directly  from ${\cal S}_{N}$. Yet, whether it is possible at all
to sample a statistically significant number of maximally compact
configurations is a moot point to consider because it involves  a
'zero' temperature sampling.

In  this paper, we present an algorithm for kinetically growing a
SAW configuration at any  given  temperature  $T  \geq  0$.  This
algorithm, called the Interacting Growth Walk (IGW) algorithm, is
able  to  generate  more  accurate data for longer walks at lower
temperatures because sample attrition is  less  severe  at  lower
temperatures.  In  fact,  on  a  square  lattice,  the walk grows
indefinitely into maximally compact configurations at $T = 0$, in
contrast to the conventional  sampling  algorithms  [9,  10].  We
demonstrate  that  our  algorithm  is  capable  of describing the
universal  behaviour  of  a  SAW  above,   at   and   below   the
$\theta$-point  in  two  dimensions.We also present a speculative
Flory thory for the IGW.

We  start the growth process by 'occupying' an arbitrarily chosen
site, ${\bf r}_{0}$, of  a  regular  $d$-dimensional  lattice  of
coordination  number  $z$  whose sites are initially 'unoccupied'
(by monomers). The first step of the walk can be made in  one  of
the  $z$  available directions, by choosing an 'unoccupied' NN of
${\bf r}_{0}$, say ${\bf r}_{1}$, at  random.  Let  the  walk  be
non-reversing  so  that  it  has a maximum of $z-1$ directions to
choose from for any further step made. Let  $\{  {\bf  r}_{j}^{m}
\mid  m=1,2,...,z_{j}  \}$  be the 'unoccupied' NNs available for
the $j^{th}$ step of the walk. If $z_{j}=0$,  the  walk  can  not
grow  further  because  it  is  geometrically  'trapped'.  It is,
therefore, discarded and a  fresh  walk  is  started  from  ${\bf
r}_{0}$.  If $z_{j} \neq 0$, the walk proceeds by choosing one of
the available sites with a probability defined as follows:

Let  $n_{NN}^{m}(j)$  be  the  number  of  nbNN  sites  of  ${\bf
r}_{j}^{m}$. Then, the probability that this site is  chosen  for
the $j^{th}$ step is given by,
\begin{equation}
p_{m}({\bf r}_{j}) \equiv  \frac {exp[-\beta n_{NN}^{m}(j) \epsilon _{0}]}
                                                        {\sum\limits_{m=1}^{z_{j}}exp[-\beta n_{NN}^{m}(j) \epsilon _{0}]}
\end{equation}
where  the  summation is over all the $z_{j}$ available sites. At
'infinite'   temperature   ($\beta   =0$),   the   local   growth
probability, $p_{m}({\bf r}_{j})$, is equal to $1/z_{j}$ and thus
the  walk  generated  will  be  the  same as the KGW. However, at
finite temperatures, the walk will prefer to  step  into  a  site
with  more  or less nbNN contacts depending on whether $\epsilon$
is  negative  or  positive.  The   probability   of   kinetically
generating  a walk configuration, ${\cal C}\equiv \{ {\bf r}_{0},
{\bf r}_{1},...,{\bf r}_{j},...\}$, is then  given  by  $P_{{\cal
C}}=  \prod_{j}  p({\bf r}_{j})$. We set $\epsilon _{0}$ equal to
$-1$ without loss of generality so that $\beta$ could  correspond
to the dimensionless temperature.
\begin{figure}
\includegraphics[width=3.25in,height=4.0in]{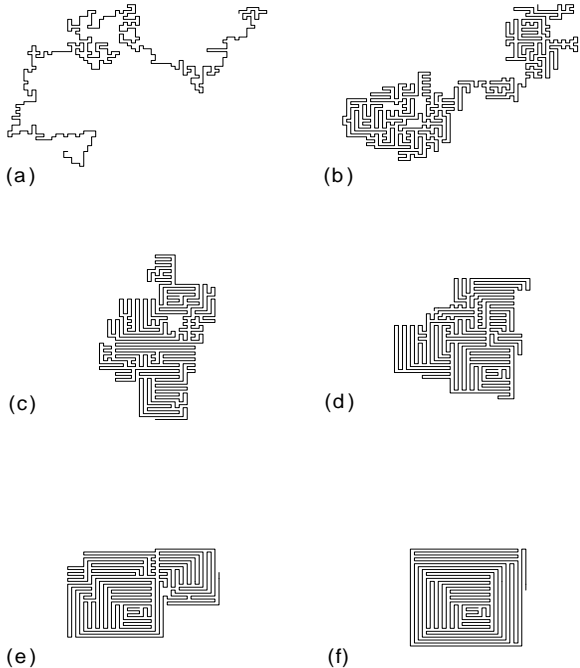}
\caption{Typical configurations of a $1000$-step walk on a square lattice
for $\beta = 0(a), 2.0(b), 3.0(c), 4.0(d), 5.0(e)$ and $300(f)$.}
\end{figure}
In   Fig.1,  we  have  shown  the  typical  configurations  of  a
$1000$-step walk on a square lattice for $\beta =  0,  2.0,  3.0,
4.0,  5.0$  and  $300$.  Evidently,  the  walk  grows into a more
compact configuration at lower temperatures, made up of  a  chain
of square blobs having 'helical' and 'sheetlike' structures.

We have generated ten million configurations of walks upto $2500$
steps  for  various  values  of  $\beta  $, and obtained the mean
square end-to-end distance, $<r^{2}(N)>$, as a simple  unweighted
average  ({\it  i.e.,}  $<r^{2}(N)>  = \sum_{\cal C} r^{2}/ {\cal
N}$,  where  the  summation  is   over   all   the   ${\cal   N}$
configurations  generated).  We  have presented $<r^{2}(N)>$ as a
function of $N$ in Fig.2.
\begin{figure}
\includegraphics[width=3.25in,height=2.25in]{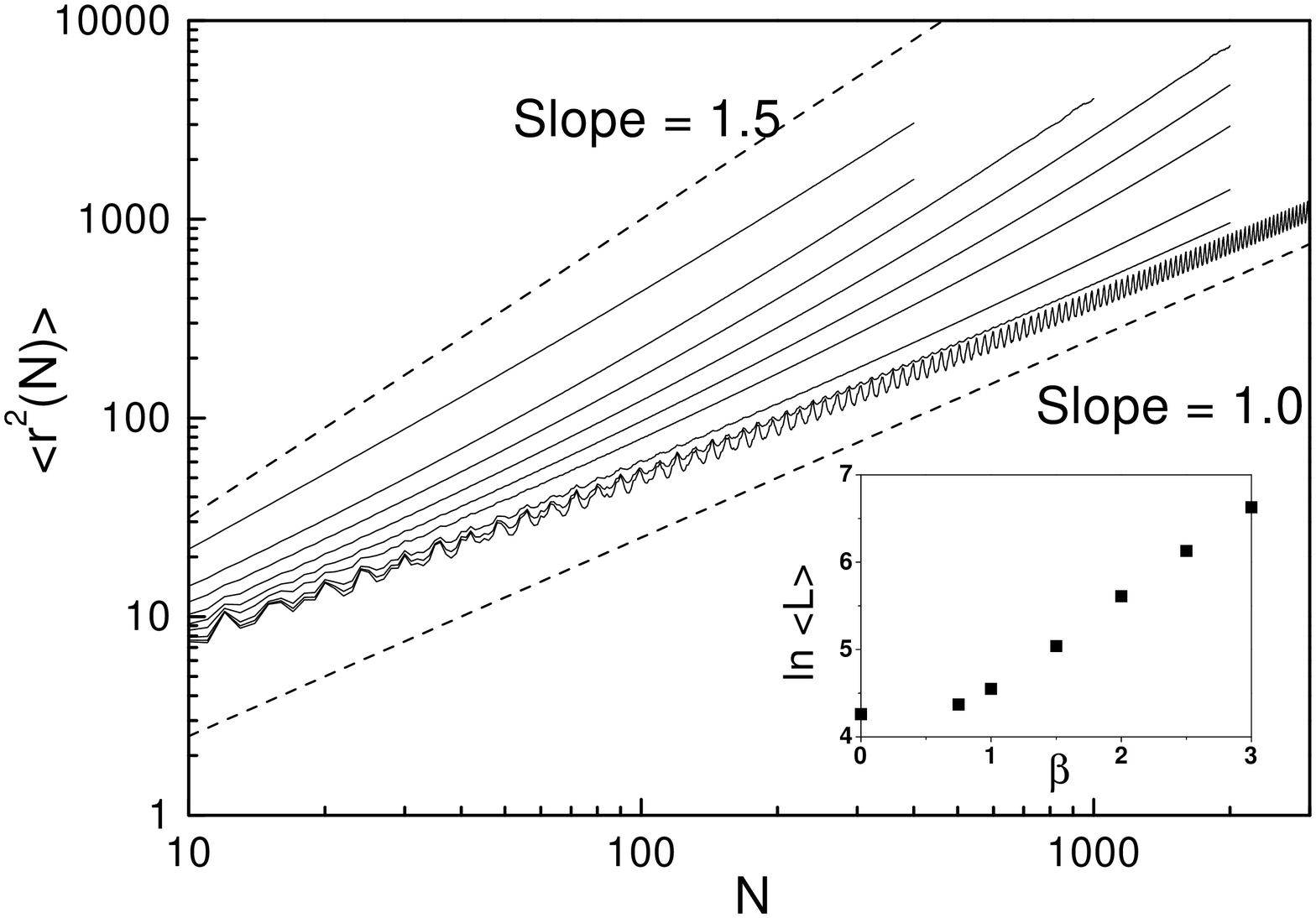}
\caption{ Log-log plot of the mean square end-to-end distance as a function of $N$ for $\beta =
0, 1.0, 1.5, 2.0, 2.5, 3.0, 4.0, 5.0$ and $300$, from top to bottom. Inset: Logarithm of the mean trapping
length, $ln <L>$ as a function of $\beta$}
\includegraphics[width=3.25in,height=2.0in]{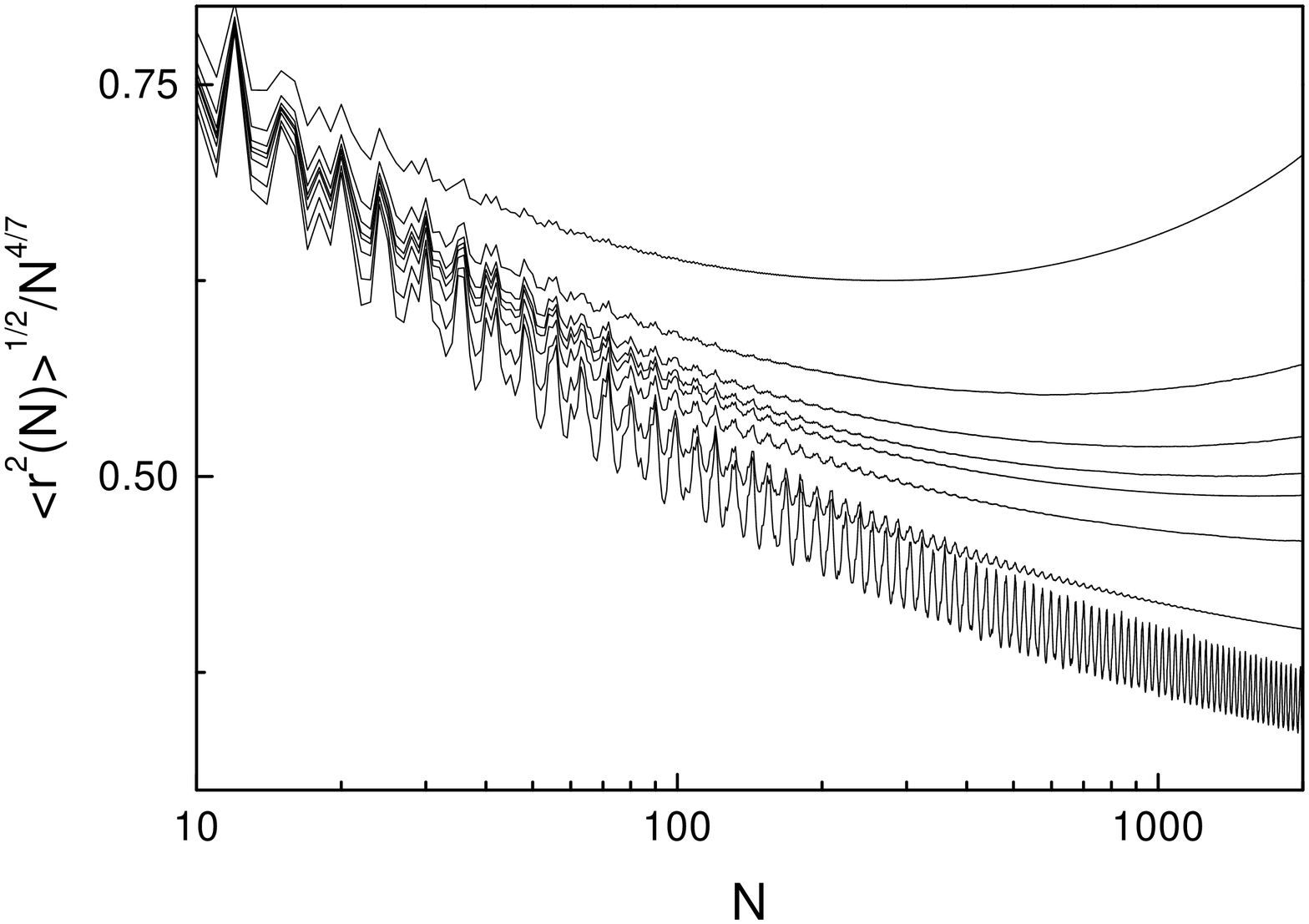}
\caption{ Semi-log plot of $<r^{2}(N)>^{1/2}/N^{4/7}$ as a function of log(N) for $\beta = 3.0, 3.5, 3.75,
3.9, 4.0, 4.25, 5.0$ and $300$, from top to bottom}
\end{figure}
Sample  attrition is the most severe problem for $\beta=0$ and it
becomes less and less severe as the value of  $\beta$  increases.
Consequently,  we  have  presented  the  data  upto  a maximum of
$N=350$ for $\beta=0$ and $N=2500$ for $\beta=300$. It  is  clear
that  the  dotted  lines with slopes $1.5$ and $1.0$ indicate the
asymptotic  behaviour  of  the  data  for  $\beta=0$  and  $\beta
\rightarrow  \infty$,  corresponding to the SAW and the collapsed
walk limits respectively. We do not know {\it a priori} whether a
collapse transition exists for our walk. We assume that it exists
and is in the same universality class as the $\theta$-point,  and
then check if our data support this assumption.

Since  it  is  known that the exponents, $\nu$ and $\gamma$, have
the exact  values  $4/7$  and  $8/7$  at  $\theta$-point  in  two
dimensions  [4],  we have plotted $<r^{2}(N)>^{1/2}/N^{4/7}$ as a
function of $log(N)$ in Fig.3. The data tend to flatten  out  for
$\beta  \sim  4$  implying  thereby  that  the  $\theta$-point is
located  near  this  value  of  $\beta$.  We  have  also  plotted
$<r^{2}(N)>/N^{8/7}$  as  a  function of $\beta$ in Fig.4 for $N=
800, 1000, 1200, 1400, 1600,  1800$  and  $2000$.  The  crossover
value  of  $\beta$  ($  \sim  4$ in our case) is expected [11] to
correspond to the $\theta$-point value.
\begin{figure}
\includegraphics[width=3.25in,height=2.0in]{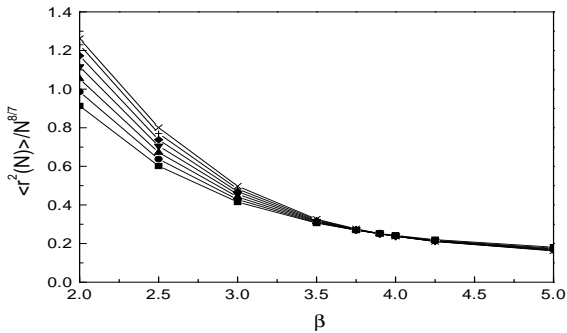}
\caption{ $<r^{2}(N)>/N^{8/7}$ as a function of $\beta$ for $N = 800$ to $2000$ in steps of $200$ from
bottom to top.}
\end{figure}

Independently,  we  have  obtained the exponent $\gamma$ from the
fraction of successful walks, $S(N) \sim  N^{\gamma-1}e^{-\lambda
N}$,  where  $\lambda$ is the attrition constant and plotted them
for six different values  of  $\beta$  in  Fig.5.  We  find  that
$\gamma$  has  a  value  ($\sim  1.13$)  close  to  the  expected
theoretical value $8/7$ for $\beta =4$.

Further evidence that it is indeed close to the $\theta$-point is
presented in Fig.6, where we have plotted the crossover exponent,
$\phi(N)$,  as  a  function  of  $1/N$  at  $\beta=4$  using  the
prescription of Grassberger and Hegger [12]. The solid line is  a
quartic  polynomial fit to the data drawn so as to guide the eye.
The extrapolated value ($0.419 \pm 0.003$) for $\phi$ is close to
the expected exact value $3/7$.
\begin{figure}
\includegraphics[width=3.25in,height=2.0in]{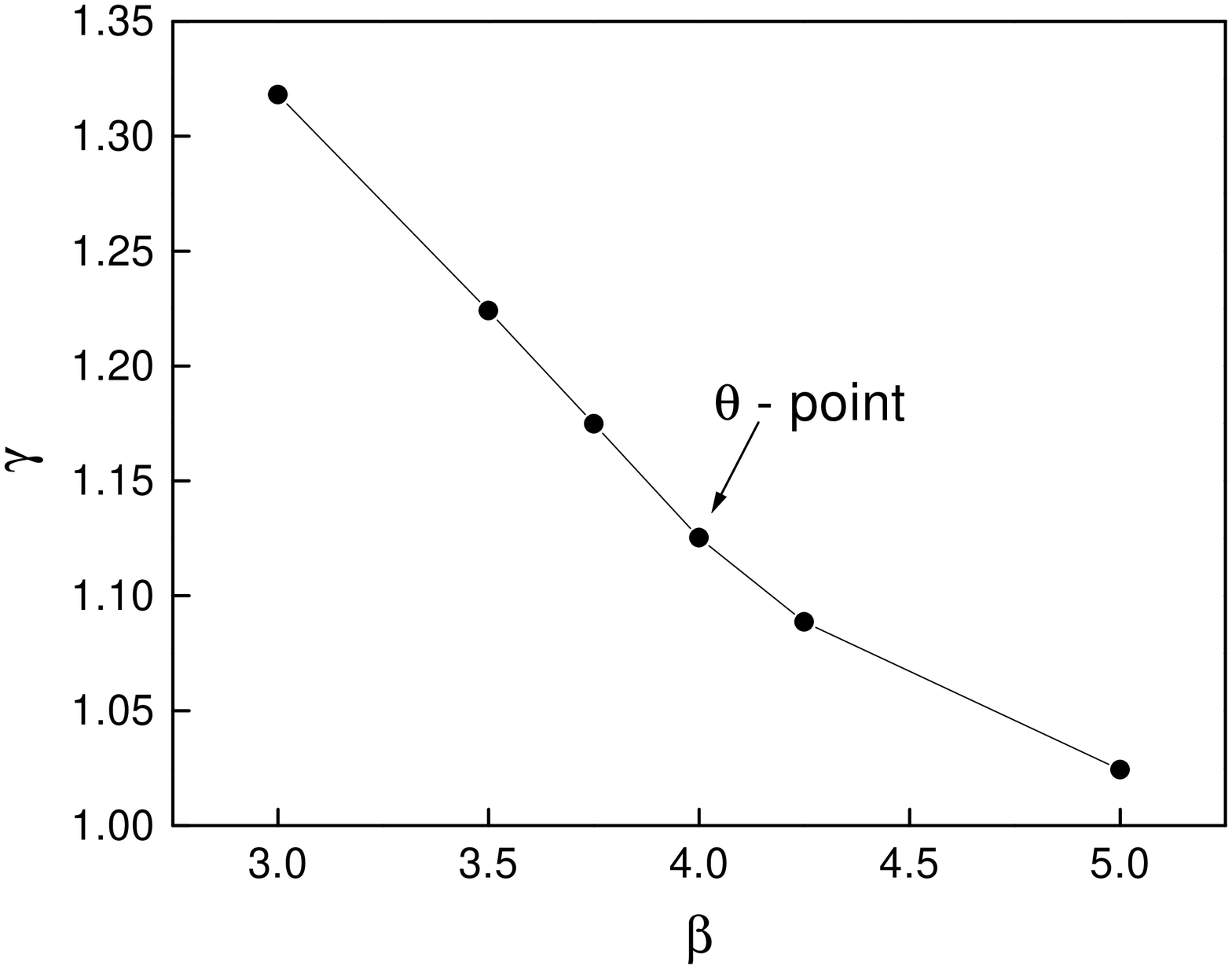}
\caption{ The exponent $\gamma$ as a function of $\beta$. Corresponding to the $\theta$-point, $\gamma$
has a value $\sim 1.13$.}
\includegraphics[width=3.25in,height=2.0in]{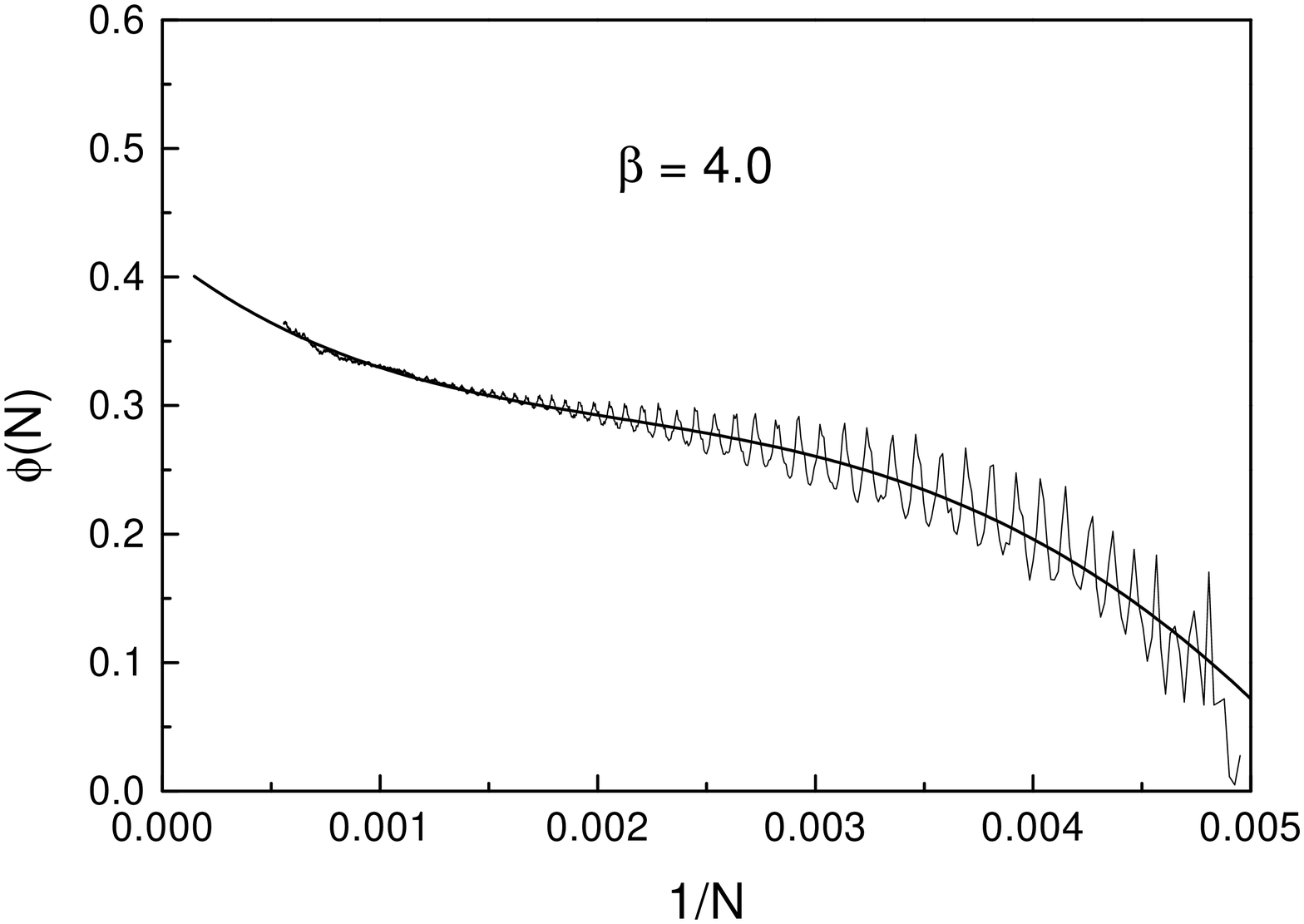}
\caption{The crossover exponent, $\phi$, as a function of $1/N$. The solid line is a quartic polynomial fit and is drawn
to guide the eye. The extrapolated value is $\sim 0.419 \pm 0.003$}
\end{figure}
All these figures put together suggest that a collapse transition
for  this  walk  exists and the corresponding dimensionless nbNN
contact energy is close to $-4$.

The   walk   configuration,   ${\cal   C}$,  having  a  total  of
$n_{NN}({\cal C}) =  \sum  _{j=1}^{N}  n_{NN}(j)$  non-bonded  NN
contacts, is grown with the probability,
\begin{equation}
P_{\cal C} = \frac {exp[-n_{NN}({\cal C})\beta \epsilon _{0}]}
                     {\prod_{j=1}^{N} \big ( \sum\limits_{m=1}^{z_{j}}exp[- n_{NN}^{m}(j)\beta \epsilon _{0}] \big )}
\end{equation}
It is possible to write the denominator, $W({\cal C})$, of the above
equation as  $e^{-n_{NN}({\cal  C})\beta  ''\epsilon  _{0}}$, where
$\beta ''$ is an effective inverse temperature. The value of $\beta ''$ will
be less (greater) than that of $\beta$ if $\epsilon _{0}$ is positive (negative or zero).
Nevertheless, ISAW algorithm can not sample the walk at an effective temperature
given by $\beta ' \equiv \mid (\beta '' -\beta)\mid $ because $\beta ''$ can only be 
estimated {\it a posteriori} on the basis of the configuration generated.
An alternative is to have a kinetic algorithm, such as what we have 
proposed in this paper, which grows a walk by sampling the available growth 
sites as per their {\it local} energies.  This is in contrast with the ISAW 
algorithm which samples fully grown and equally weighted SAW 
configurations ({\it i.e., chains}) according to their {\it total} energies.
To underline  this  basic  difference,  we  refer to our walk as the
Interacting Growth Walk (IGW).

It is appropriate at this juncture to note that the difference between
our algorithm and the  PERM algorithm (method B) of Grassberger [10]
is analogous to that between the Rosenbluth-Rosenbluth algorithm (RR) [13] 
and the KGW  [7].  Ours  is  the  finite  
temperature generalisation of the KGW, just as PERM is the finite temperature
generalisation of the RR method. There is no {\it a  priori}  reason  therefore to
expect  that  IGW  will  belong to the same universality class as
ISAW, they both being different models altogether. Yet, our  data
seem to suggest that it may well be so.

Since  the  IGW  is  equivalent  to  the  KGW in the limit $\beta
\rightarrow 0$, it is of interest to see if survival  probability
arguements  {\it  a  la}  Pietronero  [14]  could  be devised for
describing its asymptotic behaviour even if only tentatively. Let
${\cal T}_{N}$ be an ensemble of $N$-step True Self Avoiding Walk
(TSAW) [15] configurations whose end-to-end distances  are  known
to be Gaussian distributed in a space of dimension $d \geq 2$. As
we  move  along  an  arbitrarily  chosen configuration, we try to
estimate the  probability  of  surviving  self-intersections  and
geometrical   trappings.   This   involves   accounting  for  the
probability per step of encounter, $p_{E}$, and  the  probability
of trapping, $p_{T}$ which together determine the survival of the
walk.  Assuming  that  the  trapping  probability  per  step is a
constant and also that the encounter probability per step, $p_{E}
\sim \rho_{N}^{\alpha }$, where $\rho_{N}$ is the  chain  density
and $\alpha$ is the order of encounter ({\it i.e.,} the number of
nbNN   contacts),   it  has  been  shown  that  $\nu  =  (\alpha
+2)/(d\alpha +2)$ for the KGW.

The  observed  fact  that  the  IGW becomes more compact at lower
temperatures (Fig.1) implies, within the framework of  the  above
Flory-like  arguments,  that  there  should  be  an  enhancement,
$q_{E}[\rho_{N}]$,  of  the  encounter  probability   per   step,
$p_{E}$.  We expect $q_{E}[\rho_{N}]$ to increase implicitly as a
function   of   $\beta$   subject   to   the    condition    that
$q_{E}[\rho_{N}]  \rightarrow 1$ as $\beta \rightarrow 0$. On the
other hand, since the mean trapping length of IGW has been  found
to  increase  exponentially  with  $\beta$  (inset of Fig.2), the
trapping probability per step may be expected to be attenuated by
a factor proportional to  $exp(-\beta)$.  So,  if  we  assume  an
implicit    temperature    dependence,    $q_{E}[\rho_{N}]   \sim
\rho_{N}^{\beta}$,  we  can  show  that  $\nu  =  (\alpha  +\beta
+2)/[d(\alpha  +\beta)  +2]$.  While  it obviously reduces to the
Pietronero's formula in  the  limit  $\beta  \rightarrow  0$,  it
reduces  to  the  form $\nu = 1/d$ for the collapsed state in the
limit $\beta \rightarrow \infty$.  Since  first  order  encounter
($\alpha  =  1$)  is  sufficient to trap the walk, we have $\nu =
(\beta +3)/2(\beta +2)$ in two dimensions. This yields the value,
$\beta_{\theta} = 5$, corresponding to the  exact  $\theta$-point
exponent  $\nu  =  4/7$.  It may be noted that this value is fortuitously
close to our numerically estimated value. However,  in  order  to  ensure
universality of $\nu$, we should have a term proportional to  the
ratio   $\beta  /\beta  _{\theta}$  (say,  $\tilde{\beta}  \equiv
K\beta/\beta  _{\theta}$)  rather  than  $\beta$  itself  in  the
formula.  The  proportionality  constant $K$ may then be fixed by
the  $\theta$-point  value  of  $\nu$:  $K  +\alpha   =   2(1-\nu
_{\theta})/(d\nu _{\theta} -1)$, $d=1$ being a pathological case.
The fact that the first-order encounter does not trap the walk at 
$T=0$ implies that $\alpha$ also has some temperature dependence. 
Moreover, the continuous dependence of $\nu$ on $\beta$ which the
above  formula  suggests  is at variance with the fact that there
are only three universal regimes corresponding to  $\beta  <,  =$
and $>\beta_{\theta}$ respectively.  This needs further study.

We  thus  have  a  powerful  growth  algorithm for generating SAW
configurations at any given temperature, $T \geq 0$. Its strength
lies in the fact that it suffers less attrition and  is  able  to
selectively  grow  compact  configurations at lower temperatures.
Because  it  is   capable   of   generating   maximally   compact
configurations  at  zero  temperature,  it may prove to be a very
useful algorithm for studying protein folding processes. We  have
also  demonstrated explicitly in two dimensions that it correctly
describes the collapse transition of a homopolymer. Whether it is
exactly the same as the (ISAW) $\theta$-point is  an  interesting
open   question,  especially  because  the  minimum  walk  length
required to be in the asymptotic regime  increases  exponentially
with the inverse of temperature even in two dimensions.

We  are thankful to T. Prellberg and S. Bhattacharjee for helpful
comments on this work.

\medskip
\noindent $^*$glass@apsara.barc.ernet.in;
$^+$kpn@igcar.ernet.in;
$^\dagger$ramu@magnum.barc.ernet.in

\end{document}